\documentclass[10pt,letterpaper]{article} 

\usepackage{opex3}
\usepackage{epstopdf} 
\usepackage{amsmath} 
\usepackage{url}
\usepackage{cite} 
\usepackage{color}
\usepackage{bm}

\newcommand{\ket}[1]{\mid#1\rangle}
\newcommand{\vket}[1]{\mid\vec{#1}\rangle}
\newcommand{\bra}[1]{\langle#1\mid}

\newcommand{\braket}[2]{\langle#1\mid#2\rangle}
\newcommand{\expect}[1]{\langle #1 \rangle}

\begin{document}

\setlength{\tabcolsep}{15pt}

\title{Compressive Wavefront Sensing with Weak Values}

\author{Gregory A. Howland,$^{1,*}$ Daniel J. Lum$^{1}$, and John C. Howell$^{1}$}

\address{$^{1}$University of Rochester Department of Physics and
  Astronomy, 500 Wilson Blvd, Rochester, NY 14618, USA\\
}

\email{*ghowland@pas.rochester.edu} 

\begin{abstract}
  We demonstrate a wavefront sensor based on the compressive sensing,
  single-pixel camera. Using a high-resolution spatial light modulator
  (SLM) as a variable waveplate, we weakly couple an optical field's
  transverse-position and polarization degrees of freedom. By placing
  random, binary patterns on the SLM, polarization serves as a meter
  for directly measuring random projections of the real and imaginary
  components of the wavefront. Compressive sensing techniques can then
  recover the wavefront. We acquire high quality, $256\times 256$
  pixel images of the wavefront from only $10,000$
  projections. Photon-counting detectors give sub-picowatt
  sensitivity.
\end{abstract}



\section{Introduction}

High resolution wavefront sensing is extremely desirable for diverse
applications in research and industry.  Applications include measuring
atmospheric distortion for astronomy or communication
\cite{roddier:1999}, opthalmology \cite{thibos:1999}, microscopy
\cite{booth:2007}, light field imaging \cite{levoy:2006}, and adaptive
optics \cite{tyson:2010}. Fundamentally, a wavefront measurement can
be equated with measuring the quantum wavefunction
\cite{lundeen:2011}.

The most common wavefront sensor is the Shack-Hartmann sensor
\cite{platt:2001,lane:1992}, where a high-resolution CCD is placed in
the focal plane of a lenslet array. The optical power passing through
each lenslet gives a local intensity, while the displacement of each
lenslet's focal point on the CCD gives a local phase tilt. Due to the
uncertainty principle, Shack-Hartmann sensors are bandwidth-limited;
increased spatial resolution comes at the cost of phase precision. A
typical Shack-Hartmann sensor might have a spatial resolution of only
$30\times 30$ lenslets.

Recently, Lundeen et. al. used weak measurement to directly measure
the transverse wavefunction of a photonic ensemble
\cite{lundeen:2011}. By raster scanning a sliver of waveplate through
the field, they weakly couple the field's transverse-position and
polarization degrees-of-freedom. After post-selecting on the
zero-frequency component of the transverse momentum, the real and
imaginary parts of the optical field at the waveplate location are
recovered by measuring the final polarization. The measurement is
direct; detector values are directly proportional to the real or
imaginary parts of the signal. Similar experiments have used weak
measurement to trace the average trajectories of photons in the double
slit experiment \cite{kocsis:2011} and to measure the polarization
state of a qubit \cite{salvail:2013}.

The technique of Lundeen et. al. is interesting because it has no
inherent resolution limitation. However, their measurement process is
very inefficient and difficult to scale to high spatial resolution. It
requires a slow, physical raster scan of the piece of waveplate
through the detection plane. Because the polarization rotation is
small, long acquisition times are needed for sufficient
signal-to-noise ratio, particularly at low light levels. These
limitations make such a system impractical for many applications.

To solve these issues, we present a high resolution wavefront sensor
that combines Lundeen et. al.'s technique with the compressive sensing
(CS) single-pixel camera \cite{takhar:2006, baraniuk:2008}. In the
usual single-pixel camera, a digital micro-mirror device (DMD) is used
in conjunction with a single-element detector to take random, linear
projections of an intensity image. Optimization techniques are used to
recover the image from many fewer projections than pixels in the
image. For a wavefront measurement, we replace the DMD with a
twisted-nematic (TN) liquid crystal spatial light modulator
(SLM). Each SLM pixel acts as an independent, variable waveplate,
allowing us to couple transverse-position and polarization without
cumbersome scanning. By placing random, binary patterns on the SLM, we
directly measure random projections of the real and imaginary parts of
the transverse field at high resolution. The real and imaginary parts
of the field are recovered with standard, compressive sensing
algorithms.

We efficiently measure optical wavefronts at up to $256\times 256$
pixel resolution. Photon-counting detectors provide extreme low light
sensitivity, with typical detected optical power around $0.5$ pW. Our
system is compact and made of affordable, off-the-shelf
components. Like the standard single-pixel camera for intensity
imaging, our system can be easily adapted to any wavelength where
single-element detectors can be manufactured.

\section{Theory}

\begin{figure}[t]
\begin{centering}
\includegraphics[scale=0.25]{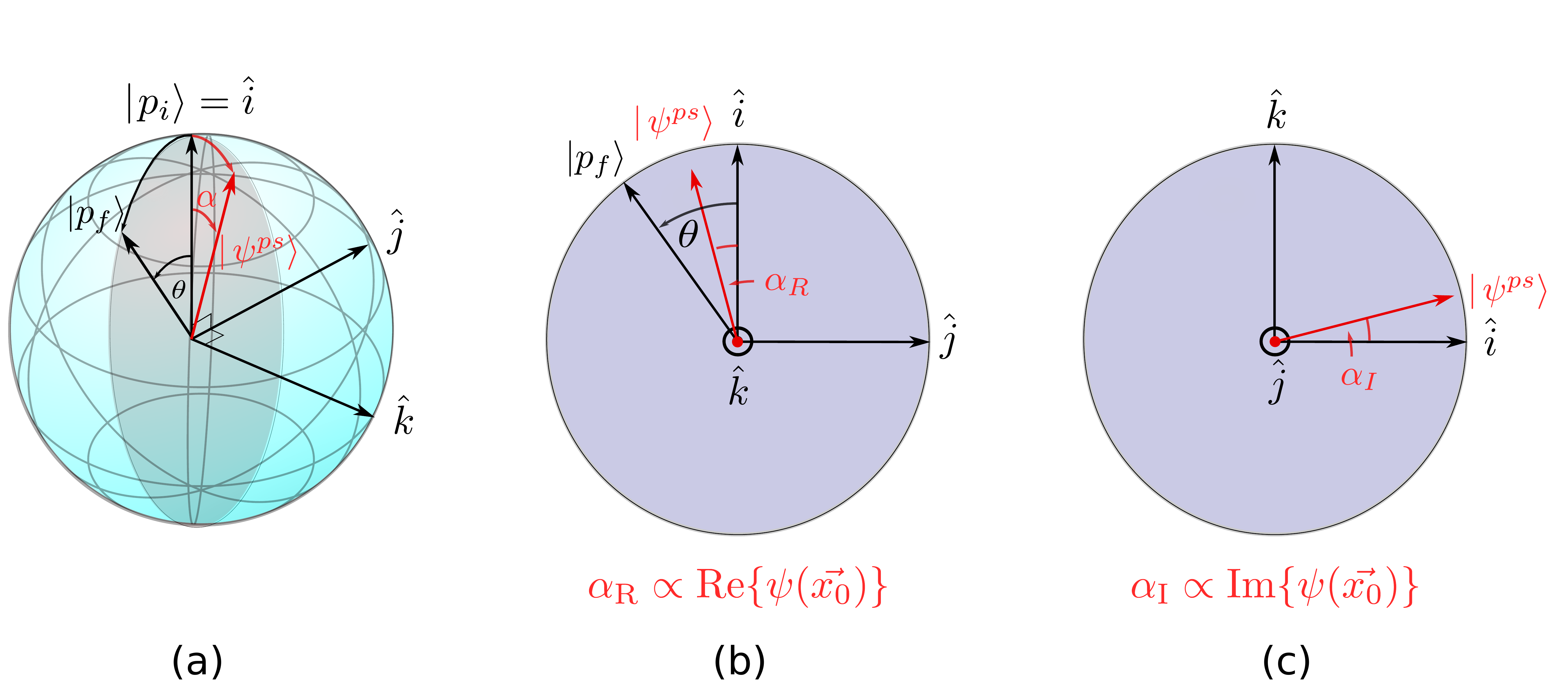}
\caption{{\bf Weak Value Wavefront Sensing on the Bloch Sphere:} A
  state with transverse field $\psi(\vec{x})$ is initially in
  polarization $\ket{p_i}$. At $\vec{x}=\vec{x_0}$, the polarization
  is rotated by small angle $\theta$ in the $(\hat{i},\hat{j})$ plane
  to state $\ket{p_f}$; this weakly measures
  $\psi(\vec{x_0})$. Following a post selection on transverse momentum
  $\vec{k}=\vec{k_0}$, the real and imaginary parts of
  $\psi(\vec{x_0})$ are mapped to a polarization rotation of the
  post-selected state $\ket{\psi^{ps}}$. The real part generates a
  rotation $\alpha_{\text{R}}$ in the $(\hat{i},\hat{j})$ plane (b)
  and the imaginary part generates a rotation $\alpha_{\text{I}}$ in
  the $(\hat{k},\hat{j})$ plane (c). The weak measurement of
  $\psi(\vec{x_0})$ is read-out by measuring $\expect{\hat{\sigma}_j}$
  and $\expect{\hat{\sigma}_k}$ respectively.}
\label{fig:bloch}
\end{centering}
\end{figure}

\subsection{Weak Measurement}

The weak value, originally presented by Aharonov, Albert, and Vaidman
as a ``new kind of value for a quantum variable''
\cite{aharonov:1988}, arises from averaging weak measurements on pre-
and post-selected systems. Originally a theoretical curiosity, weak
measurement has seen resurgent interest as it has turned out to be
very useful, particularly for precision measurement
\cite{hosten:2008,dixon:2009,jordan:2014} and investigation of
fundamental phenomena, such as Hardy's paradox \cite{lundeen:2009}.

In a weak measurement, a system of interest is investigated by very
weakly coupling it to a measuring device. The system is first prepared
in an initial state $\ket{\psi}$. An observable of interest $\hat{A}$
is then coupled to an ancillary meter system by a weak interaction or
perturbation. Finally, the system is post-selected (projected) into
final state $\ket{f}$. The weak measurement is read-out by a measuring
device for the meter. In the limit of a very weak interaction, the
measuring device's pointer is shifted by the weak value
\begin{equation}
  A_w = \frac{\bra{f}\hat{A}\ket{\psi}}{\braket{f}{\psi}}.
\end{equation}

Note that $A_w$ can be complex valued. The real part of $A_w$ is
interpreted as a shift in the pointer's position; the imaginary part
of $A_w$ is interpreted as a shift in the pointer's momentum
\cite{dressel:2012}. Because the meter's position and momentum are
observables, this enables direct measurement of complex values. For an
introduction to experimental weak measurement and weak values, see
Refs. \cite{dressel:2013,jordan:2014}.

\subsection{ Wavefront Sensing with Weak Values}
\label{sec:wvsens}

We wish to use weak measurement to directly measure an optical field
$\psi(\vec{x})$, where $\vec{x} =(x,y)$ are transverse spatial
coordinates. In keeping with traditional presentation of weak
measurement, we use a quantum formalism where $\psi(\vec{x})$ is
treated as a probability amplitude distribution and polarization is
represented on the Bloch sphere. Note that the system can still be
understood classically, replacing the wavefunction with the transverse
electric field and Bloch sphere with the Poincar\'{e} sphere
respectively.

Consider a weak measurement of position at $\vec{x}=\vec{x_0}$
followed by a post-selection of momentum $\vec{k}=\vec{k_0}$, where
$\psi(\vec{x})$ and $\tilde{\psi}(\vec{k})$ are Fourier transform
pairs. The corresponding weak value is
\begin{equation}
  A_w(\vec{x_0}) = \frac{\braket{\vec{k_0}}{\vec{x_0}}\braket{\vec{x_0}}{\psi}}{\braket{\vec{k_0}}{\psi}} = \frac{e^{ik_0x_0} \psi(\vec{x_0})}{\tilde{\psi}(\vec{k_0})}.
\end{equation}
$A_w(\vec{x_0})$ is directly proportional to value of the field at
position $\vec{x}={\vec{x_0}}$ up to a linear phase.

To obtain $A_w(\vec{x_0})$, the field at $\vec{x}=\vec{x_0}$ must be
weakly coupled to a meter system. The polarization degree of freedom
is a convenient meter because it can be easily manipulated and
measured. Let $\psi(\vec{x})$ be initially polarized in polarization
state $\ket{p_i}$. The full initial state is
\begin{equation}
  \ket{\psi} = \int \vec{dx}\psi(\vec{x})\vket{x}\ket{p_i}.
\end{equation}
At location $\vec{x}=\vec{x_0}$, the polarization is changed from
$\ket{p_i}$ to a nearby polarization $\ket{p_f}$
(Fig. \ref{fig:bloch}a-b). At location $\vec{x_0}$, the state is
\begin{equation}
  \psi(\vec{x_0})\vket{x_0}\ket{p_f} = \psi(\vec{x_0})e^{-i \hat{\sigma_k} \theta/2}\vket{x_0}\ket{p_i},
\end{equation}
where the transformation from $\ket{p_i}$ to $\ket{p_f}$ is expressed
as a rotation on the Bloch sphere by angle $\theta$ about unit vector
$\hat{k}$.  This is visualized on the Bloch sphere in
Fig. \ref{fig:bloch}. Unit vectors $\hat{i}$, $\hat{j}$, and $\hat{k}$
form a right handed coordinate system on the Bloch sphere,
where $\hat{i}$ points along $\ket{p_i}$, $\hat{j}$ is the orthogonal
unit vector in the plane defined by $\ket{p_i}$ and $\ket{p_f}$, and
$\hat{k} = \hat{i} \times \hat{j}$ (Fig. \ref{fig:bloch}a-c). These
unit vectors have corresponding Pauli operators $\hat{\sigma}_i$,
$\hat{\sigma}_j$, and $\hat{\sigma}_k$. Note that such a coordinate
system can be defined for any two polarization states $\ket{p_i}$ and
$\ket{p_f}$.

For a weak interaction, $\theta$ is small. A first order expansion at
$\vec{x_0}$ yields
\begin{equation}
  \psi(\vec{x_0})\vket{x_0}\ket{p_f} = \psi(\vec{x_0})(1-i\hat{\sigma_k}\theta/2)\vket{x_0}\ket{p_i}.
\end{equation}
The full state is therefore
\begin{equation}
  \ket{\psi} = \int \vec{dx}\psi(\vec{x})\vket{x}\ket{p_i} - \psi(\vec{x_0})i\hat{\sigma_k}\theta/2 \vket{x_0}\ket{p_i}.
\end{equation}
Consider post-selection on a single transverse-momentum
$\vec{k}=\vec{k_0}$. The post-selected state $\ket{\psi_{ps}}$ no
longer has position dependence and is given by
\begin{equation}
  \ket{\psi^{ps}} =\braket{\vec{k_0}}{\psi}=\tilde{\psi}(\vec{k_0})\ket{p_i}-e^{\vec{k_0}\cdot \vec{x_0}}\psi(\vec{x_0})i\hat{\sigma_k}\theta/2
  \ket{p_i},
\end{equation}
where $\tilde{\psi}(\vec{k})$ is the Fourier transform of $\psi(\vec{x})$. 

Factoring out $\tilde{\psi}(\vec{k_0})$ and re-exponentiating, we find
\begin{equation}
  \ket{\psi^{ps}}=\tilde{\psi}(\vec{k_0})e^{-ie^{i\vec{k_0}\cdot\vec{x_0}}\frac{\psi(\vec{x_0})}{2\tilde{\psi}(\vec{k_0})}\hat{\sigma_k}\theta}\ket{p_i} = \tilde{\psi}(\vec{k_0})e^{-iA_w(\vec{x_0})\hat{\sigma_k}\theta/2}\ket{p_i}
\end{equation}

The post-selected polarization state is simply a rotated version of
the initial polarization, where the rotation is proportional to
$\psi(\vec{x_0})$ (Fig. \ref{fig:bloch}a-c). The real part of
$\psi(\vec{x_0})$ generates a rotation $\alpha_R$ in the
$\hat{i},\hat{j}$ plane (Fig. \ref{fig:bloch}b). The imaginary part of
$\psi(\vec{x_0})$ generates a rotation $\alpha_I$ in the
$\hat{i},\hat{k}$ plane (Fig. \ref{fig:bloch}c).

The values are therefore measured by taking expected values of Pauli operators
$\hat{\sigma_j}$ and $\hat{\sigma_k}$
\begin{align}
\bra{\psi^{ps}}\hat{\sigma_j}\ket{\psi^{ps}} &\propto \text{Re}\{\psi(\vec{x_0})\}\\
\bra{\psi^{ps}}\hat{\sigma_k}\ket{\psi^{ps}} &\propto \text{Im}\{\psi(\vec{x_0})\}.
\end{align}

\subsection{Random Projections of the Wavefront}
\label{sec:proj}

Rather than only measuring the wavefunction at a single location
$\psi(\vec{x} = \vec{x_0})$, consider instead a weak measurement of an
operator $\hat{f}_i$ which takes a random, binary projection of
$\psi(\vec{x})$, where $\ket{f_i}$ is
\begin{equation}
  \ket{f_i} = \int d\vec{x} f_i(\vec{x})\ket{\vec{x_i}}.
\end{equation}
The filter function $f_i(\vec{x})$ consists of a pixelized, random
binary pattern, where pixels in the pattern take on values of 1 or -1
with equal probability.

The weak measurement of $\hat{f}_i$, given initial state
$\psi(\vec{x})$ and post-selected state $\ket{\vec{k_0}}$, is
therefore
\begin{equation}
  A_i = \frac{\braket{\vec{k_0}}{f_i}\braket{f_i}{\psi}}{\braket{\vec{k_0}}{\psi}} =
  \frac{\braket{\vec{k_0}}{f_i}Y_i}{\tilde{\psi}(\vec{k_0})},
   \label{eq:aproj}
\end{equation}
where $Y_i$ is the inner product between $\psi(\vec{x})$ and $f_i(\vec{x})$
\begin{equation}
  Y_i = \int d\vec{x} f_i(\vec{x})\psi(\vec{x}).
\end{equation}
It is convenient to choose $\vec{k_0}=(0,0)$ to discard the linear
phase factor $\braket{\vec{k_0}}{f_i}$ in Eq. \ref{eq:aproj}.

To perform the weak measurement, we again couple transverse-position
to polarization. Unlike the previous case, all of $\psi(\vec{x})$ will
now receive a small polarization rotation about $\hat{k}$ of angle
$\theta$ for $f_i(\vec{x})=1$ and $-\theta$ for $f_i(\vec{x})=-1$;.

Performing an identical derivation to section \ref{sec:wvsens}, we
find a post-selected polarization state
\begin{equation}
  \ket{\psi^{ps}_{i}} = \tilde{\psi}(\vec{k_0})e^{-i\frac{Y_i}{2\psi(\vec{k0})}\hat{\sigma}_k \theta}\ket{p_i}.
\end{equation}
The effective polarization rotation is now proportional to the projection of $\psi(\vec{x})$ onto $f_i(\vec{x})$, $Y_i$. Again, taking expectation values of $\hat{\sigma}_j$ and $\hat{\sigma}_k$ yields the real and imaginary parts of $Y_i$,
\begin{align}
  \bra{\psi^{ps}_{i}}\hat{\sigma}_j\ket{\psi^{ps}_i} &\propto Y^{\text{Re}}_i \\
  \bra{\psi^{ps}_{i}}\hat{\sigma}_k\ket{\psi^{ps}_i} &\propto
  Y^{\text{Im}}_i.
\end{align}
Therefore, weak measurement allows us to directly measure random,
binary projections of a transverse field $\psi(\vec{x})$.

\subsection{Compressive Sensing}
\label{sec:cs}

The random, binary projections of section \ref{sec:proj} are the type
of measurements used in Compressive
Sensing\cite{donoho:2006}. Compressive sensing is a measurement
technique that compresses a signal during measurement, rather than
after, to dramatically decrease the requisite number of measurements.

In compressive sensing, one seeks to recover a compressible,
$N$-dimensional signal $X$ from $M<<N$ measurements. The signal $X$ is
linearly sampled by a $M\times N$ sensing matrix $F$ to produce an
$M$-dimensional vector of measurements $Y$
\begin{equation}
  Y = FX + \Phi,
\label{eq:meas}
\end{equation}
where $\Phi$ is an $M$-dimensional noise vector. Measurement vectors
(rows of F) are often random, binary vectors, so each measurement
$Y_i$ is a random projection of $X$ \cite{candes:2008}.

Because $M<<N$, the system of equations in Eq. \ref{eq:meas} is
under-determined; there are many possible signals consistent with the
measurements. CS posits that the correct $X$ is the one which is
sparsest (has the fewest non-zero elements) when compressed. This $X$
is found by solving a regularized, least squares objective function
\begin{equation}
  \min_X \frac{\mu}{2}||Y-FX||^{2}_{2}+g(X).  
  \label{eq:objective}
\end{equation}
The first penalty is small when $X$ is consistent with $Y$. The second
penalty, $g(X)$, is small when $X$ is compressible. A common $g(X)$
for imaging is the Total Variation (TV) of $X$
\begin{equation}
  \text{TV}(X) = \sum_{\text{adj. }i,j}X_i-X_j,
\end{equation}
where indices $i$ and $j$ run over all pairs of adjacent pixels. TV
therefore leverages compressibility in the gradient of $X$. In this
case, Eq. \ref{eq:objective} is referred to as Total Variation
Minimization\cite{chambolle:1997}.

Remarkably, exact recovery of a $k$-sparse (only $k$ significant
elements when compressed) $X$ is possible from only $M\propto
k\log(\frac{N}{k})$ measurements, which can be as low as a few percent
of $N$ \cite{candes:2006}.

The most well-known example of Compressive Sensing is the single-pixel
camera for intensity imaging \cite{baraniuk:2008}. A scene of interest
$X$ is imaged onto a digital micro-mirror device (DMD), an array of
mirrors which can be individually oriented towards or away from a
single-element detector. A series of $M$ random binary patterns are
placed on the DMD, each corresponding to a row of sensing matrix
$F$. The total optical power striking the detector for the
$i^{\text{th}}$ pattern gives the projection of $X$ onto $F_i$, the
$i^{\text{th}}$ measurement $Y_i$. Solving Eq. \ref{eq:objective}
recovers the image.

CS has found many applications including magnetic resonance imaging
\cite{lustig:2007}, radio astronomy \cite{bobin:2008} and quantum
entanglement characterization \cite{gross:2010, howland:2013}. For a
thorough introduction to Compressive Sensing, see
Refs. \cite{candes:2008:2,romberg:2008}.

\subsection{Compressive Wavefront Sensing}

\begin{figure}[t]
  \begin{centering}
    \includegraphics[scale=0.4]{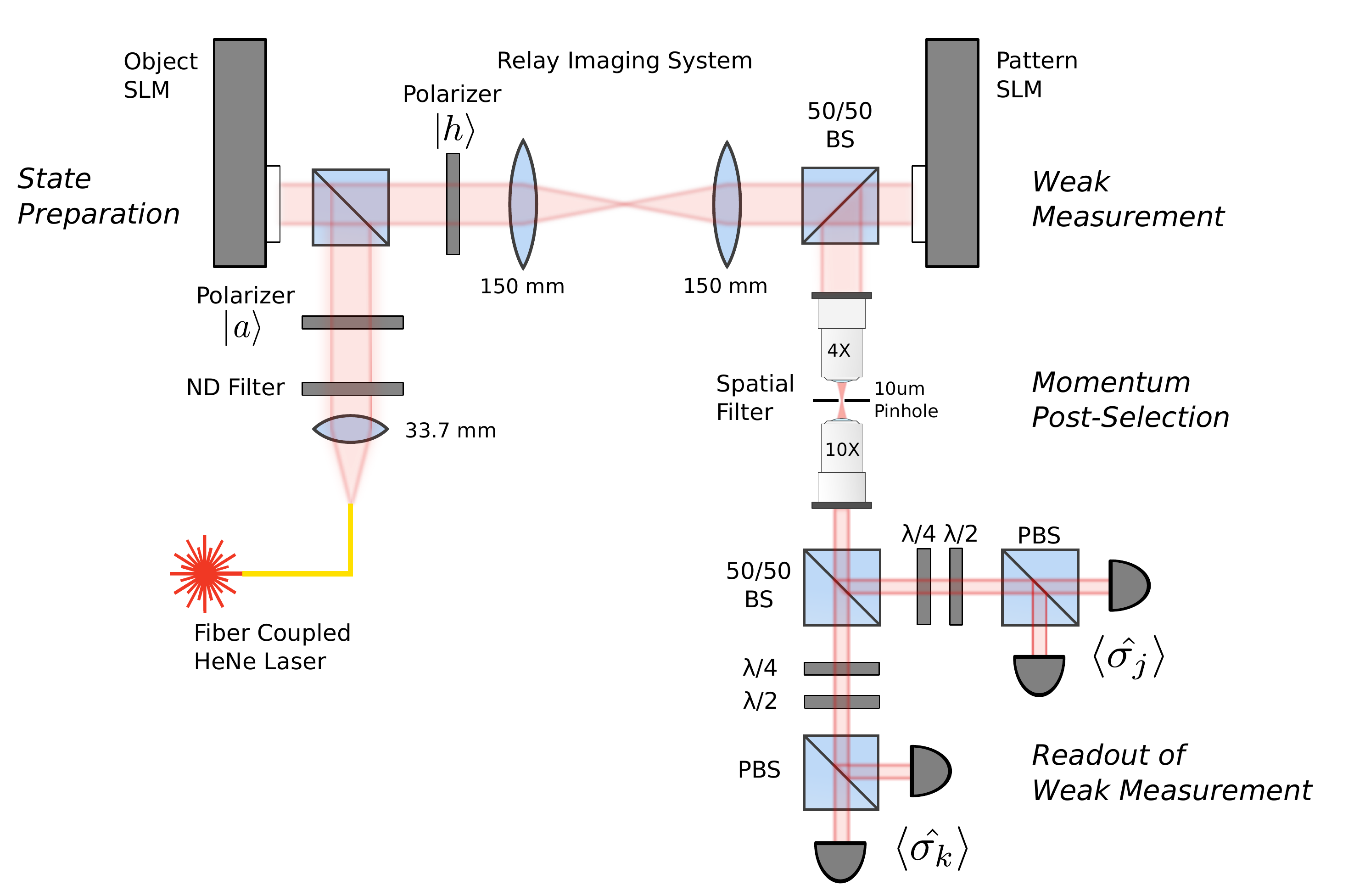}
    \caption{{\bf Experimental Setup for Wavefront Sensing:} An input
      field is prepared by illuminating an object SLM with a
      collimated, attenuated HeNe laser. The input beam is polarized
      in $\ket{a}$ to produce a nearly pure phase object with some
      intensity coupling. A 4F imaging system reproduces the field on
      a pattern SLM, with a polarizer setting the initial polarization
      to horizontal ($\ket{p_i}=\ket{h}$). A sequence of $M$ random,
      binary patterns are placed on the SLM; pattern pixels with value
      $1$ have their polarization rotated a small amount to
      $\ket{p_f}$ This constitutes a weak measurement of the
      projection of the input field onto the random pattern. A spatial
      filter performs the $\vec{k}=0$ momentum
      post-selection. Polarization analyzers take expected values of
      $\hat{\sigma}_j$ and $\hat{\sigma}_k$, which are proportional to
      the real and imaginary parts of the projection. }
    \label{fig:exp}
  \end{centering}
\end{figure}

We now have the building blocks to implement a single-pixel,
compressive wavefront sensor. We first use weak measurement to take a
series of $M$ random projections of the real and imaginary parts of
the wavefront (sec \ref{sec:proj}). We then use compressive sensing
optimization algorithms to recover the real and imaginary parts of the
field.

The random projections require a device that can perform
position-dependent polarization rotations. A twisted-nematic, liquid
crystal spatial light modulator (SLM) is capable of this by acting as
a variable waveplate. For our SLM, each pixel performs the operation
\begin{equation}
  \hat{T}(\vec{x_0}) = \ket{d}\bra{d} + exp(i\Phi(I_{\text{SLM}}(\vec{x_0})))\ket{a}\bra{a},
\end{equation}
where $I_{\text{SLM}}(\vec{x_0})$ is the SLM intensity at location
$\vec{x_0}$ and $\ket{d}$ and $\ket{a}$ refer to diagonal and
anti-diagonal polarizations respectively. For a typical $8$-bit SLM
driven over a VGA video port, $I_{\text{SLM}}$ takes on integer values
of $0-255$. Because SLM pixels retard $\ket{a}$ and not $\ket{d}$, a
non-zero SLM intensity will rotate the polarization of any input state
that is not purely $\ket{a}$ or $\ket{d}$.

Let N-dimensional signal vector $X = X^{\text{Re}}+iX^{\text{Im}}$ be
a one-dimensional reshaping of $\psi(\vec{x_0})$, discretized to the
SLM resolution. Let $F$ be a $M\times N$, random, binary sensing
matrix whose elements take values $1$ or $-1$ with equal
probability. Each row of $F$ corresponds to a pattern placed on the
SLM, where pixels with value $1$ rotate the field's polarization by
$\theta$ and pixels with value $-1$ rotate the field's polarization by
$-\theta$.

The random projection of $X$ onto each pattern is weakly measured as
in section \ref{sec:proj} to produce an $M$-dimensional measurement
vector 
\begin{equation}
  Y = Y^{\text{Re}}+iY^{\text{Im}} = FX.
\end{equation}

The real and imaginary parts of $X$ are recovered by solving
Eq. \ref{eq:objective}.

\section{Experiment}

\begin{figure}[t]
\begin{centering}
\includegraphics[scale=0.5]{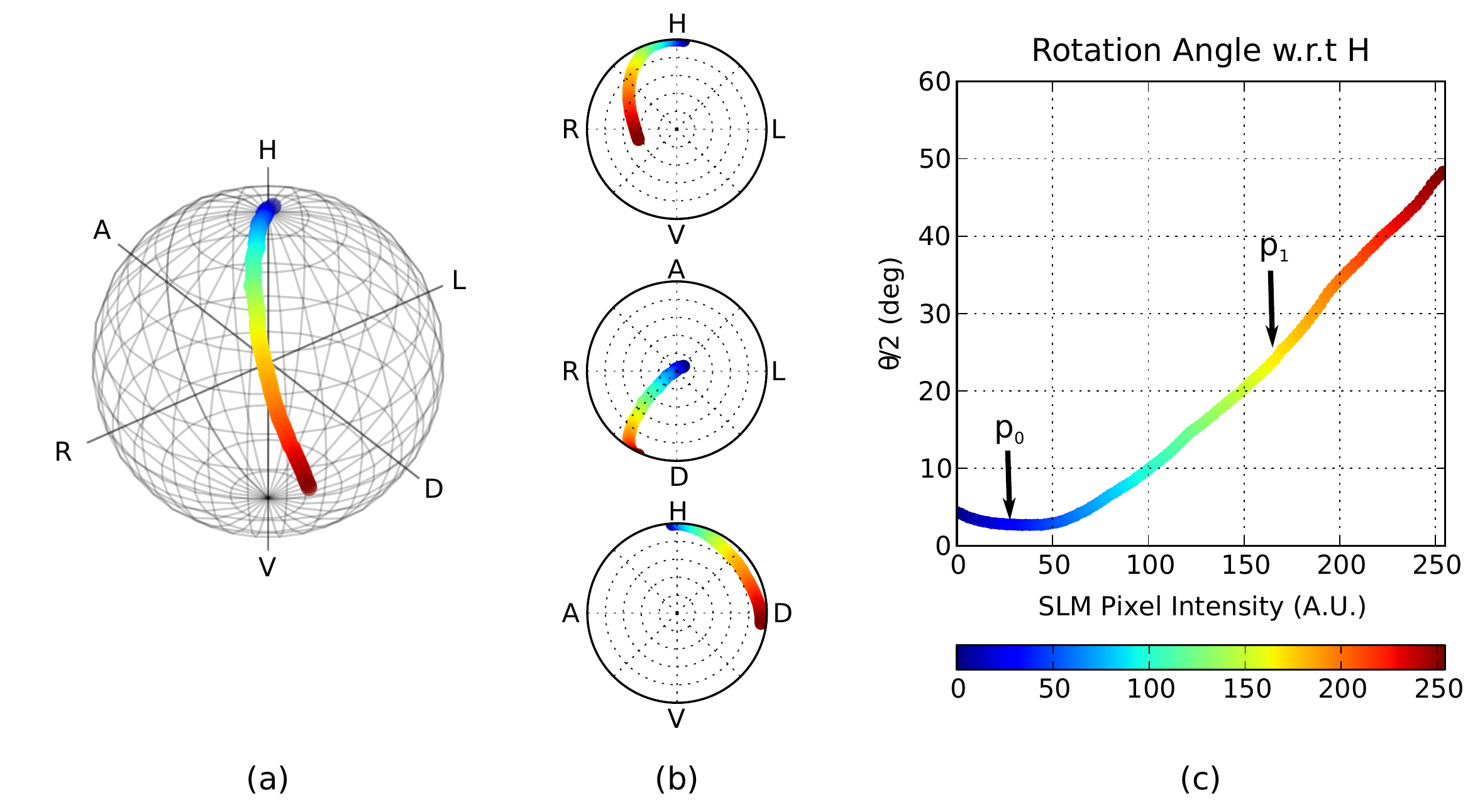}
\caption{{\bf SLM Polarization Calibration:} The polarization rotation
  performed by the SLM on a horizontally polarized input state on the
  Bloch sphere is given in (a) and (b). (c) gives the angle $\theta/2$
  between the initial horizontal state $\ket{p_i}$ and output state
  $\ket{p_f}$. Point $p_0$ corresponds to a minimal rotation less than
  $1$ degree. Point $p_1$ is the SLM intensity and corresponding angle
  used for the rotated state $\ket{p_f}$, approximately $25$ degrees.}
\label{fig:pcal}
\end{centering}
\end{figure}

\subsection{Experimental Apparatus}

The experimental apparatus is given in Fig. \ref{fig:exp}. A beam from
a fiber-coupled, HeNe laser is collimated by a $33.7$ mm focal length
lens. A neutral density filter reduces the optical power to the
single-photon regime, approximately $0.5$ pW. An input field to be
measured is prepared with a $\ket{a}$ oriented polarizer and an object
SLM (Cambridge Correlators SDE1024). Because the SLM retards
$\ket{a}$, an image placed on it will produce a nearly pure phase
image with a small amount of amplitude coupling.

A 4F imaging system reproduces the field from the object SLM onto a
pattern SLM which is used to perform the weak measurement. A
horizontal ($\ket{h}$) polarizer prepares the initial polarization
state $\ket{p_i}$. A sequence of $M$ random, binary patterns are
placed on the SLM, executing the weak measurement. The patterns
consist of randomly-permuted rows of a Hadamard matrix. This
dramatically improves reconstruction algorithm speeds as repeated
calculations of $FX$ can be performed by a fast transform
\cite{li:2011}.

Following the weak measurement, a spatial filter consisting of a 4X
microscope objective and a $10$ $\mu$m pinhole performs the
$\vec{k}=0$ post-selection. The Gaussian beam exiting the pinhole is
collimated with a $10$X objective and directed to a pair of
polarization analyzers each consisting of a half-waveplate,
quarter-waveplate, and a polarizing beamsplitter. The half- and
quarter-waveplates are oriented to measures either
$\expect{\hat{\sigma}_j}$ or $\expect{\hat{\sigma}_k}$ for the
respective real and imaginary projections. The detectors are large
area, photon-counting photomultiplier modules (Horiba TBX-900C).

The difference in the count rate between each polarization analyzer's
outputs for each pattern make up $Y^{\text{Re}}$ and
$Y^{\text{Im}}$. To solve Eq. \ref{eq:objective}, we use the TVAL3
solver \cite{li:2009}.

\begin{figure}[b]
\begin{centering}
\includegraphics[scale=0.4]{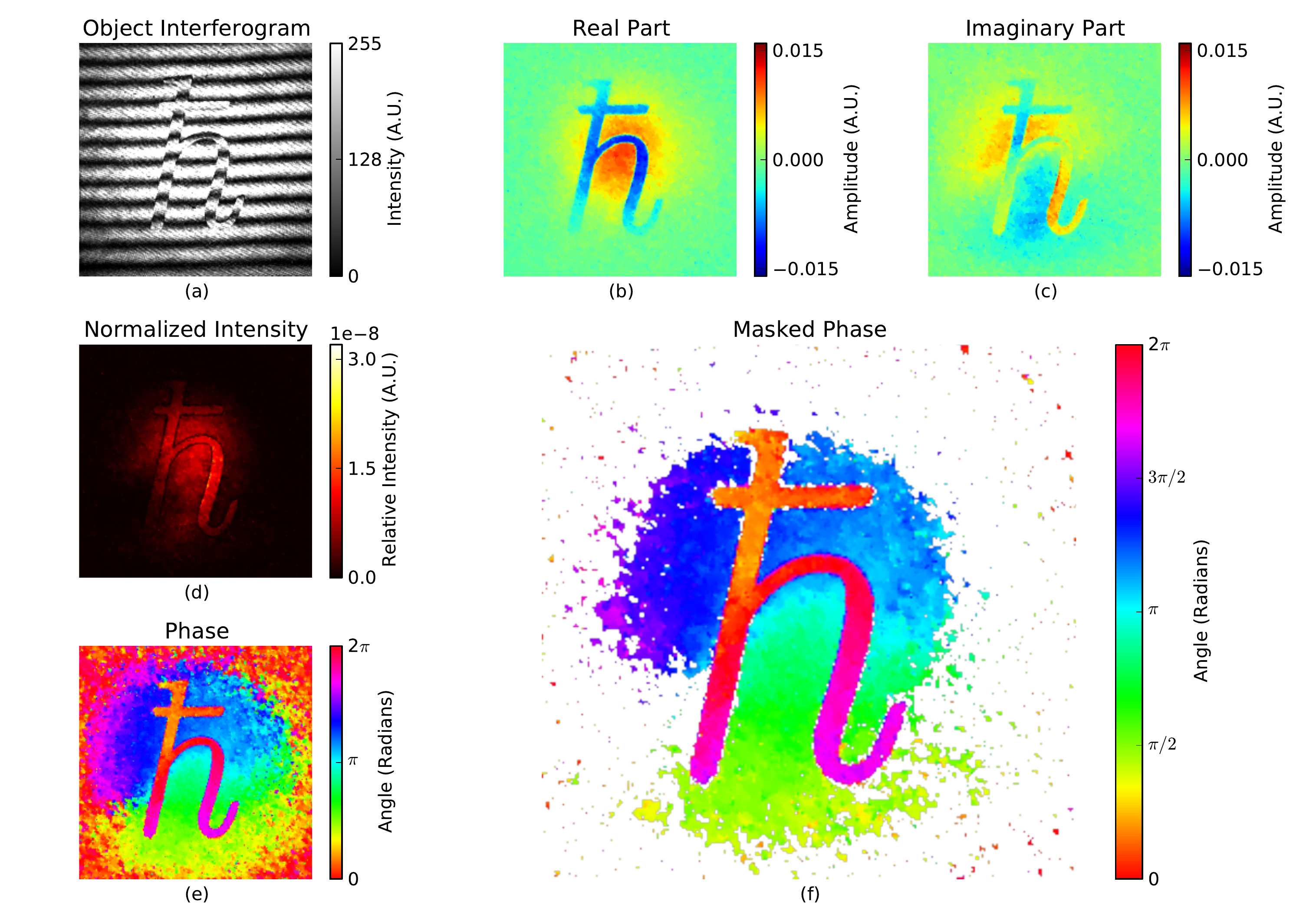}
\caption{{\bf $\bm{256\times 256}$ pixel $\bm{\hbar}$ Character
    Wavefront:} An interferogram of the object field (a) taken with an
  $8$-bit CCD camera confirms a near-phase-only input wavefront. The
  CS-reconstructed real and imaginary parts are shown in (b,c), where
  the wavefront is normalized to unit total intensity. From the real
  and imaginary parts, we find intensity (d) and phase (e) images. A
  masked phase image (f) removes all intensities below $5\times
  10^{-10}$, where it is not meaningful to asign a phase. Only
  $M=0.15N=10,000$ random projections are needed for a high quality
  reconstruction.}
\label{fig:hbar}
\end{centering}
\end{figure}

\begin{figure}[t]
\begin{centering}
\includegraphics[scale=0.5]{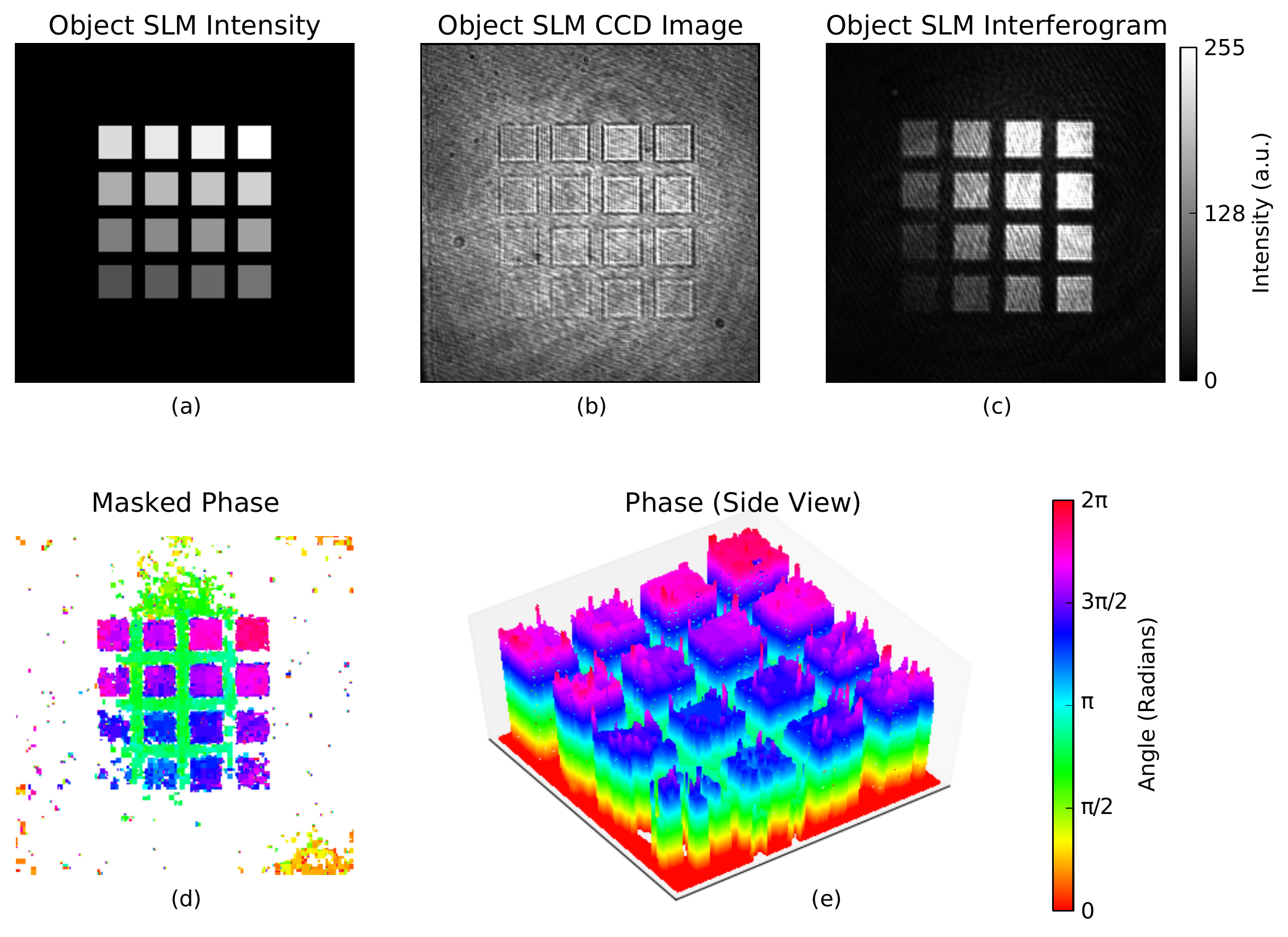}
\caption{{\bf Phase Grid Test Pattern:} A $16$ square grid of
  increasing SLM intensity (a) was placed on the object SLM, which
  converts it to a phase image. A CCD image of the object SLM (b)
  shows near-uniform intensity, while a dark-port interferogram (c)
  shows the increasing phase of each square.  The phase angle for a
  $256\times 256$ reconstruction is given in (d-e) for $M=10000$. The
  increasing phase of each square can be seen riding the illuminating
  beam's Gaussian phase profile. }
\label{fig:squares}
\end{centering}
\end{figure}

\subsection{SLM Polarization Rotation}

A calibration of the SLM polarization rotation as a function of SLM
intensity is given in Fig. \ref{fig:pcal}. To calibrate, we illuminate
the SLM with a horizontally polarized beam. A uniform intensity was
placed on the SLM, and the output polarization was analyzed. The SLM
intensity is then scanned through its $8$-bit range
$0-255$. Fig. \ref{fig:pcal}a gives the output polarization on the
Bloch sphere, with projections into principle planes given in
Fig. \ref{fig:pcal}(b). Fig. \ref{fig:pcal}(c) gives the angle
$\theta/2$ between the output polarization and horizontal.

The selected polarization rotation $\theta/2$ for pixels with value
$1$ was chosen to be $25$ degrees with a corresponding
$I_{\text{SLM}}=160$. Because the SLM is not able to simultaneously
perform a $-\theta/2$ rotation, two measurements must be taken for
each SLM pattern $f_i$. First, all pixels with value $1$ set to
$I_{\text{SLM}}=160$, and all pixels with value $-1$ are set to
$I_{\text{SLM}}=20$. The latter experience a polarization rotation
less than $1$ degree; effectively no rotation. Then, the pattern is
inverted; pixels with value -1 are set to $I_{\text{SLM}}=160$ and
pixels with value $1$ are set to $I_{\text{SLM}}=20$. Subtracting the
value for the second situation from the first achieves the desired
result of a positive $\theta$ rotation for ``1'' pixels and a
$-\theta$ rotation for ``-1'' pixels.

\section{Results}

The recovered field for a $256\times 256$ pixel phase $\hbar$
character is given in Fig. \ref{fig:hbar}. Fig. \ref{fig:hbar}a shows
a camera image of the interference between the object field and
reference beam. The image is predominantly a phase-only image with a
small amount of amplitude coupling, particularly at the edges of
$\hbar$ character. Fig. \ref{fig:hbar}b-c show the reconstructed real
and imaginary parts of field, while Fig. \ref{fig:hbar}d-e give its
intensity and phase-angle. The reconstructed field is normalized to
have unit total intensity. Fig. \ref{fig:hbar}f gives a masked phase
image, where values with negligible intensity below $5\times 10^{-10}$
are colored white.

The recovered field has near uniform intensity up to the width of the
illumination beam, but a strong variation in phase corresponding to
the $\hbar$ character. Only $M = .15N = 10000$ random projections were
used to recover a high quality image.

Fig. \ref{fig:squares} lists the results for a grid of $16$ phase
squares. Fig. \ref{fig:squares}(a) is the image placed on the object
SLM, where each square increases in intensity from the bottom left to
the top right from $I_{\text{SLM}} = 30$ to $I_{\text{SLM}}=240$. The
SLM converts this image to a phase image. These values were chosen
because the polarization rotation appears roughly linear on this range
(see Fig. \ref{fig:pcal}).  Fig. \ref{fig:squares}(b) is a CCD image
of the object SLM that shows that the object field has near-uniform
intensity. Fig. \ref{fig:squares}(c) gives a dark port interferogram
demonstrating the validity of our phase-square reconstruction, where
the bottom left square has an almost negligible phase shift with the
background while the top right square is nearly $\pi$ out of phase.

TVAL3 was initially used to solve Eq. \ref{eq:objective}. However,
because of the smoothly varying Gaussian wave-front in the background
and additional shot-noise in the measurement process, TVAL3 found a
sparse solution in the gradient of the image that did not have a
minimized least-squares. Simply stated, TVAL3 found the edges within
the image without assigning proper values to them.

As shown in previous work, the solver Gradient Projection for Sparse
Reconstruction (GPSR) can be used as a least-squares fitting tool on
the largest components of the signal returned by TVAL3 in a Haar
wavelet basis \cite{figueiredo:2007,howland:2013:2}.  A mask was
obtained in the pixel basis by neglecting all values less than $4\%$
of the maximum intensity from TVAL3 while least-squares was performed
in the wavelet basis with GPSR. The minimized components were then
transformed back into the pixel basis and multiplied by our mask to do
away with the unwanted artifacts. Fig. \ref{fig:squares}(d) and
\ref{fig:squares}(e) are the $256\times 256$ pixel final results. The
reconstructed values accurately reflect the trend set in
Fig. \ref{fig:squares}(c).

\section{Conclusion}

We have demonstrated a wavefront sensor that combines the efficiency
and flexibility of compressive sensing with weak measurement. Because
our technique directly measures random projections of the real and
imaginary parts of a transverse optical field, it is not subject to
space-bandwidth product limitations. Compressive sensing makes the
technique practical; the transverse field can be acquired from many
fewer measurements than pixels ($M<<N$) with only single-element
detectors and without scanning. We anticipate that our technique will
be a valuable and useful addition to the field of wavefront sensing
and adaptive optics.

\section{Acknowledgments}

The figures in this manuscript were produced in part with Matplotlib
\cite{hunter:2007}. We gratefully acknowledge support from AFOSR grant
FA9550-13-1-0019, DARPA DSO InPho grant W911NF-10-1-0404, and ARO
grant W911NF-12-1-0263.


\begin{thebibliography}{10}
\newcommand{\enquote}[1]{``#1''}

\bibitem{roddier:1999}
F.~Roddier, \emph{Adaptive optics in astronomy} (Cambridge university press,
  1999).

\bibitem{thibos:1999}
L.~N. Thibos and X.~Hong, \enquote{Clinical applications of the
  {S}hack-{H}artmann aberrometer,} Optometry \& Vision Science \textbf{76},
  817--825 (1999).

\bibitem{booth:2007}
M.~J. Booth, \enquote{Adaptive optics in microscopy,} Philosophical
  Transactions of the Royal Society A: Mathematical, Physical and Engineering
  Sciences \textbf{365}, 2829--2843 (2007).

\bibitem{levoy:2006}
M.~Levoy, \enquote{Light fields and computational imaging,} IEEE Computer
  \textbf{39}, 46--55 (2006).

\bibitem{tyson:2010}
R.~Tyson, \emph{Principles of adaptive optics} (CRC Press, 2010).

\bibitem{lundeen:2011}
J.~S. Lundeen, B.~Sutherland, A.~Patel, C.~Stewart, and C.~Bamber,
  \enquote{Direct measurement of the quantum wavefunction,} Nature
  \textbf{474}, 188--191 (2011).

\bibitem{platt:2001}
B.~C. Platt and R.~Shack, \enquote{History and principles of {S}hack-{H}artmann
  wavefront sensing,} Journal of Refractive Surgery \textbf{17}, S573--S577
  (2001).

\bibitem{lane:1992}
R.~Lane and M.~Tallon, \enquote{Wave-front reconstruction using a
  {S}hack-{H}artmann sensor,} Applied optics \textbf{31}, 6902--6908 (1992).

\bibitem{kocsis:2011}
S.~Kocsis, B.~Braverman, S.~Ravets, M.~J. Stevens, R.~P. Mirin, L.~K. Shalm,
  and A.~M. Steinberg, \enquote{Observing the average trajectories of single
  photons in a two-slit interferometer,} Science \textbf{332}, 1170--1173
  (2011).

\bibitem{salvail:2013}
J.~Z. Salvail, M.~Agnew, A.~S. Johnson, E.~Bolduc, J.~Leach, and R.~W. Boyd,
  \enquote{Full characterization of polarization states of light via direct
  measurement,} Nature Photonics \textbf{7}, 316--321 (2013).

\bibitem{takhar:2006}
D.~Takhar, J.~N. Laska, M.~B. Wakin, M.~F. Duarte, D.~Baron, S.~Sarvotham,
  K.~F. Kelly, and R.~G. Baraniuk, \enquote{A new compressive imaging camera
  architecture using optical-domain compression,} in \enquote{Electronic
  Imaging 2006,}  (International Society for Optics and Photonics, 2006), pp.
  606509--606509.

\bibitem{baraniuk:2008}
R.~G. Baraniuk, \enquote{Single-pixel imaging via compressive sampling,} IEEE
  Signal Processing Magazine  (2008).

\bibitem{aharonov:1988}
Y.~Aharonov, D.~Z. Albert, and L.~Vaidman, \enquote{How the result of a
  measurement of a component of the spin of a spin-1/2 particle can turn out to
  be 100,} Physical review letters \textbf{60}, 1351 (1988).

\bibitem{hosten:2008}
O.~Hosten and P.~Kwiat, \enquote{Observation of the spin hall effect of light
  via weak measurements,} Science \textbf{319}, 787--790 (2008).

\bibitem{dixon:2009}
P.~B. Dixon, D.~J. Starling, A.~N. Jordan, and J.~C. Howell,
  \enquote{Ultrasensitive beam deflection measurement via interferometric weak
  value amplification,} Physical review letters \textbf{102}, 173601 (2009).

\bibitem{jordan:2014}
A.~N. Jordan, J.~Mart\'inez-Rinc\'on, and J.~C. Howell, \enquote{Technical
  advantages for weak-value amplification: When less is more,} Phys. Rev. X
  \textbf{4}, 011031 (2014).

\bibitem{lundeen:2009}
J.~Lundeen and A.~Steinberg, \enquote{Experimental joint weak measurement on a
  photon pair as a probe of hardy’s paradox,} Physical review letters
  \textbf{102}, 020404 (2009).

\bibitem{dressel:2012}
J.~Dressel and A.~Jordan, \enquote{Significance of the imaginary part of the
  weak value,} Physical Review A \textbf{85}, 012107 (2012).

\bibitem{dressel:2013}
J.~Dressel, M.~Malik, F.~M. Miatto, A.~N. Jordan, and R.~W. Boyd,
  \enquote{Understanding quantum weak values: Basics and applications,} arXiv
  preprint arXiv:1305.7154  (2013).

\bibitem{donoho:2006}
D.~L. Donoho, \enquote{Compressed sensing,} Information Theory, IEEE
  Transactions on \textbf{52}, 1289--1306 (2006).

\bibitem{candes:2008}
E.~J. Candes, \enquote{The restricted isometry property and its implications
  for compressed sensing,} Comptes Rendus Mathematique \textbf{346}, 589--592
  (2008).

\bibitem{chambolle:1997}
A.~Chambolle and P.-L. Lions, \enquote{Image recovery via total variation
  minimization and related problems,} Numerische Mathematik \textbf{76},
  167--188 (1997).

\bibitem{candes:2006}
E.~J. Candes and T.~Tao, \enquote{Near-optimal signal recovery from random
  projections: Universal encoding strategies?} Information Theory, IEEE
  Transactions on \textbf{52}, 5406--5425 (2006).

\bibitem{lustig:2007}
M.~Lustig, D.~Donoho, and J.~M. Pauly, \enquote{Sparse {MRI}: The application
  of compressed sensing for rapid {MR} imaging,} Magnetic resonance in medicine
  \textbf{58}, 1182--1195 (2007).

\bibitem{bobin:2008}
J.~Bobin, J.-L. Starck, and R.~Ottensamer, \enquote{Compressed sensing in
  astronomy,} Selected Topics in Signal Processing, IEEE Journal of \textbf{2},
  718--726 (2008).

\bibitem{gross:2010}
D.~Gross, Y.-K. Liu, S.~T. Flammia, S.~Becker, and J.~Eisert, \enquote{Quantum
  state tomography via compressed sensing,} Physical review letters
  \textbf{105}, 150401 (2010).

\bibitem{howland:2013}
G.~A. Howland and J.~C. Howell, \enquote{Efficient high-dimensional
  entanglement imaging with a compressive-sensing double-pixel camera,}
  Physical Review X \textbf{3}, 011013 (2013).

\bibitem{candes:2008:2}
E.~J. Cand{\`e}s and M.~B. Wakin, \enquote{An introduction to compressive
  sampling,} Signal Processing Magazine, IEEE \textbf{25}, 21--30 (2008).

\bibitem{romberg:2008}
J.~Romberg, \enquote{Imaging via compressive sampling [introduction to
  compressive sampling and recovery via convex programming],} IEEE Signal
  Processing Magazine \textbf{25}, 14--20 (2008).

\bibitem{li:2011}
C.~Li, \enquote{Compressive sensing for {3D} data processing tasks:
  applications, models and algorithms,} Ph.D. thesis, Rice University (2011).

\bibitem{li:2009}
C.~Li, W.~Yin, and Y.~Zhang, \enquote{User’s guide for {TVAL3}: {TV}
  minimization by augmented lagrangian and alternating direction algorithms,}
  CAAM Report  (2009).

\bibitem{figueiredo:2007}
M.~A. Figueiredo, R.~D. Nowak, and S.~J. Wright, \enquote{Gradient projection
  for sparse reconstruction: Application to compressed sensing and other
  inverse problems,} Selected Topics in Signal Processing, IEEE Journal of
  \textbf{1}, 586--597 (2007).

\bibitem{howland:2013:2}
G.~A. Howland, D.~J. Lum, M.~R. Ware, and J.~C. Howell, \enquote{Photon
  counting compressive depth mapping,} Optics express \textbf{21}, 23822--23837
  (2013).

\bibitem{hunter:2007}
J.~D. Hunter, \enquote{Matplotlib: A {2D} graphics environment,} Computing In
  Science \& Engineering \textbf{9}, 90--95 (2007).

\end{thebibliography}
\end{document}